# The Fair Distribution of Power to Electric Vehicles: An Alternative to Pricing

Yingjie Zhou[*,†,††], *Member*, *IEEE*, Nicholas Maxemchuk[†,#], *Fellow*, *IEEE*, Xiangying Qian[†], and Chen Wang[†]
[*]College of Computer Science, Sichuan University, Chengdu, China
[†]Department of Electrical Engineering, Columbia University, New York, USA
[#]Institute IMDEA Networks, Madrid, Spain
[††]University of Electronic Science and Technology of China, Chengdu, China
Email: yjzhou09@gmail.com, nick@ee.columbia.edu

***Abstract*—As the popularity of electric vehicles increases, the demand for more power can increase more rapidly than our ability to install additional generating capacity. In the long term we expect that the supply and demand will become balanced. However, in the interim the rate at which electric vehicles can be deployed will depend on our ability to charge these vehicles without inconveniencing their owners. In this paper, we investigate using fairness mechanisms to distribute power to electric vehicles on a smart grid. We assume that during peak demand there is insufficient power to charge all the vehicles simultaneously. In each five minute interval of time we select a subset of the vehicles to charge, based upon information about the vehicles.**

**We evaluate the selection mechanisms using published data on the current demand for electric power as a function of time of day, current driving habits for commuting, and the current rates at which electric vehicles can be charged on home outlets. We found that conventional selection strategies, such as first-come-first-served or round robin, may delay a significant fraction of the vehicles by more than two hours, even when the total available power over the course of a day is two or three times the power required by the vehicles. However, a selection mechanism that minimizes the maximum delay can reduce the delays to a few minutes, even when the capacity available for charging electric vehicles exceeds their requirements by as little as 5%.**

## I. INTRODUCTION

An emphasis on green technologies and the price of gasoline is causing the number of electric vehicles to increase rapidly. However, the time to approve and construct new generating facilities is longer than the time required to build and distribute electric vehicles. Therefore, it is likely that there will be times when the generating capacity will not be sufficient to meet the power demand created by electric vehicles. To allow the use of electric vehicles to increase as rapidly as possible, it's necessary to find techniques to manage electric vehicle charging so that their owners are not inconvenienced.

An important difference between charging electrical vehicles and other loads is that the demand need not be serviced immediately. The vehicle is likely to be plugged in for a longer period of time than it takes to recharge the battery, particularly when vehicles are charged at home. At present, there are several proposals to shift power consumption from peak hours by charging less for power during non-peak hours [1-4].

Rather than using pricing to encourage vehicles to shift their load to non-peak hours, we use information collected on a smart grid to select the rate at which different vehicles receive charge. The techniques for distributing the available power among vehicles are based on fairness mechanisms that have been used to allocate resources in communication networks [8]. We simulate a system that uses the selection mechanisms and evaluate the mechanisms based on real world measurements.

The selection mechanisms are implemented by inserting an on/off switch between the outlet and the battery charger for each electric vehicle. We do not control the rate at which a vehicle receives charge, but how often it receives charge. At any point in time, the power available to charge electric vehicles is the power that can be generated minus the power supplied to other loads. The available power determines the number of electric vehicles that can be charged simultaneously. When the number of vehicles that are charging exceeds the number that can be charged, the power company selects the vehicles that receive charge by operating their on/off switch. Time is divided into five-minute intervals, and during each interval the power company selects the subset of the vehicles that receive charge. The fraction of the intervals that the switch is actuated is the rate at which the vehicles receive charge. This strategy is straightforward to implement and should operate with any battery charging system that can be plugged into a standard outlet.

Our objective is to compare rules that the power company can use to select the vehicles that receive charge. The rules that we described in the paper are:

1) Round robin (RR), that cycles through the list of vehicles requesting charge, so that all of the vehicles receive charge in the same fraction of five minute intervals;

2) First come first serve (FCFS), that charges the vehicles in the order that they arrive;

3) First depart first serve (FDFS), that charges the vehicles in the order that they are expected to leave;

4) Min-max Energy Requirement (MinmaxER), that first charges the vehicles that require the most energy to reach their destination;

5) And, Min-max Delay (MinmaxDT), that first charges the vehicles with the longest delays beyond their expected departure time.

The rules are described in detail in section III. The first two selection rules can be implemented using information that can be measured, the number of vehicles that are connected to their chargers, the time when the vehicles were plugged into their chargers, and their battery levels. The third rule requires the expected departure time of the vehicles. The first three rules can be improved by considering the amount of charge that the vehicles require to reach their destinations, rather than charging the vehicles until their batteries are fully charged. We use the required charge in the evaluations for a fairer comparison with the fourth and fifth rules. The delays in the first three rules increase if we do not use the expected driving distances. The fourth and fifth rules require the expected driving distances. And, the fifth rule also requires the expected departure time of the vehicles, as did the third rule.

The expected departure time and expected driving distance cannot be measured, and must be obtained by estimating them from the past use of the vehicle, or by requiring the user to enter values. The main thrust of this work is not to determine the best way to obtain these values, but to determine whether or not they are worth obtaining. Our conclusion is that the fifth rule outperforms the other rules by such a large margin, that we should determine the best way to obtain these values. In section V, the conclusion, we present one possible way to obtain these values.

When a vehicle has not acquired sufficient charge to reach its destination at the time that the user would like to leave, the user is delayed until he has acquired sufficient charge. The selection rules are evaluated based upon the fraction of the vehicles that are forced to leave late, and the average time that those vehicles are delayed.

Our objective is to delay users as infrequently as possible and by as little time as possible, in the most energy constrained conditions. The metric that we use to quantify the degree to which the power network is energy constrained is: (energy available to charge vehicles in a 24 hour period) over (the energy required to charge all of the vehicles during the same 24 hour period). We assume that the ratio is greater than 1. We may not have sufficient energy to charge all of the vehicles that are currently plugged in, but over the course of the day the vehicles can be charged. Vehicles are delayed, but they are not denied charging. We compare the selection mechanisms by determining the fraction of vehicles that are delayed and the time that they are delayed as a function of the constraint ratio.

The selection rules make it unnecessary to have rolling blackouts or brownouts, the current approach for dealing with power shortages, to cope with the added requirements of electric vehicles. They are intuitively fairer than pricing mechanisms that auction off power during the time of the day when the load exceeds the generating capacity. Pricing favors the wealthier consumers, who are better able to pay. In addition, the selection rules are more precise than pricing and can make better use of all of the available power. Pricing mechanisms encourage users to delay charging until a time when there is more power available, but do not control the exact number of users who delay charging. If more users than necessary delay their charging, not all of the available power is used. The selection mechanisms control the exact number of users who are delayed, and also determines which of the users are delayed. The power demand is trimmed with a scalpel rather than a butcher's knife.

The real world measurements that we use to evaluate the charging mechanisms are presented in section II. In section III, we present the five mechanisms for allocating power, and in section IV we compare them based on the real world data. We present our conclusions in section V.

## II. Real World Measurements

In this section, we present the real world measurements that we use to simulate our selection mechanisms.

In our model, most users arrive at home and charge their vehicles after work. The 2009 American Community Survey Report [5], provides the distribution for the average number of workers leaving home to go to work as a function of the time of day. We shifted the time axis by 10 hours, *i.e.*, 8 a.m. corresponds to 6 p.m., to get the average rate at which vehicles arrive for charging. The shift is based on an average work day of 8 hours and an two additional hours for commuting and errands. The arrivals follow a Poisson distribution with a rate $\lambda$ that changes as a function of time of day. The Poisson distribution is appropriate when the vehicles operate independently.

The total number of electric vehicles that are simulated in the system is scaled to around 1500. This is appropriate for a medium-sized city in the U.S. based on President Obama's goal of putting one million electric vehicles on the road by 2015. We assume that 80 large-size cities will have about 5000 electric vehicles each and 400 medium-sized cities will have about 1500 electric vehicles each, to reach the goal of 1,000,000.

Vehicles are connected to their chargers for an average of 14 hours with a standard deviation of 4 hours. The charging time for a specific vehicle is selected from a Normal distribution that is truncated below 6 hours and above 22 hours (a probability of 4.55%). The desired departure time for a vehicle is its arrival time plus the time that it is connected to its charger.

The power that is used by loads other than electric vehicles changes during a twenty-four hour day [7]. We assume that the maximum power that can be generated by the power company is fixed, and that the power that is available for electric vehicles at any time during the day is the generating capacity minus the power used for other loads. It is fortuitous that most electric vehicles are charged during the night, when the smallest amount of power is used by other loads and the largest amount of power is available to charge electric vehicles.

The commuting distances that we use in the simulations are based on the Omnibus Household Survey by the US Department of Transportation [6]. The survey presents the distribution of one-way commuting distances, which we have doubled to determine the round trip commuting distances. We have fitted an Exponential distribution to the survey data, and truncated the commute distances at 70 miles. It is unlikely that a person who commutes more than 70 miles per day will buy a vehicle that can only travel 100 miles between recharging. When we pick a commuting distance, we add 20 miles for other driving during the day and 10 miles for emergencies, to determine the charge that a vehicle must acquire before leaving the charging station. In this set of simulations we assume that vehicles are charged to the level required to reach their destinations, rather than a full battery. When a vehicle arrives at the charging station, the battery level is uniformly distributed between 0 and 30 miles, depending on the discretionary driving on that day.

In the simulations, vehicles are plugged into a standard home socket, 110 volt, 15 amps. Most of the current generation of electric vehicles, including the Nissan LEAF, FORD 2012 Focus Electric, BMW Mini E, THINK City, Mitsuibishi iMeiv and SMART, have similar power requirements. They require 28kwh to travel approximately 100 miles. If a vehicle arrives with an empty battery, it must be plugged into a standard home socket for about 17 hours, or 200 5-minute charging intervals, to receive enough charge to travel 100 miles. For each hour that an electric vehicle is plugged into a standard home socket, it can only travel about 6 miles. In a practical system the charging rate will most likely be slower because it is recommended that the continuous current on a 15 amp circuit remain below 13 amps. We will also consider using higher voltage and higher amperage circuits, similar to those that are used in homes for electric clothes dryers and whole house air conditioners. The higher the voltage and current, the faster electric vehicles can be charged, however, these circuits will probably require adding new circuits to a home, which is contrary to current advertizing.

## III. FAIRNESS MECHANISMS

In section III.A, we formulate the problem of allocating the power that is available to charge electric vehicles and present the metrics that we use to evaluate the selection mechanisms. The distribution grid is smart enough to: 1) determine the amount of power that is currently available to charge electric vehicles; 2) count the number of vehicles that require charge; 3) record the time when a vehicle arrives for charging; 4) monitor a vehicle's battery level; 5) obtain estimates on expected departure times and driving distances; and, 6) control a switch at each electric vehicle that determines whether or not the vehicle receives charge in each five minute interval.

In the remaining sub-sections, we describe the selection rules that we will compare. The rules use different subsets of the information.

### A. Problem Formulation

The fair allocation of power to charge electric vehicles is a resource allocation problem that is similar to allocating flows in a communications network [9]. Several of the techniques that we describe for electric vehicles are similar to those that have been used in networks. The resource that we are allocating is the power that is available to charge electric vehicles, and the objective is to provide each vehicle with sufficient charge for its expected driving distance before its expected departure time.

There are significant differences between the current problem and resource allocation in networks. 1) In networks when we cannot provide every user with all of the capacity that they request, we provide each user with a fraction of their request. In our charging problem, all of the vehicles must acquire their requested charge before they leave. The metrics that we use to evaluate charging techniques are the fraction of the vehicles that are delayed and the amount of time that they are delayed. 2) In a network the capacity that is allocated to the users remains constant. In vehicle charging, the power that is available for vehicles changes continuously as the power required by other loads varies.

The total power available for charging electric vehicles during a 24 hour day, *TPA*, is the total power that can be generated minus the power that is consumed by other loads. The effect of the time varying requirements of other loads depends on the fraction of the total generating capacity that is required by these loads. In the simulations presented in this paper, we assume that the maximum amount of power required for other loads is 80% of the generating capacity.

The total power required to charge all of the electric vehicles during a 24 hour day is *TPR*, and the supply to demand ratio for electric vehicles is:

$$SDR = TPA / TPR \quad (1)$$

If $SDR<1$, not all of the vehicles that arrive during the course of a day can be charged, and the delays will grow indefinitely. In the simulations we note that when $SDR>3$, all the power distribution mechanisms work reasonably well. However, when $1\leq SDR\leq 3$, the power distribution mechanisms have significantly different delay characteristics. We use two metrics to compare the delays, the fraction of the vehicles that are delayed, *FOD*, and the average amount of delay that the delayed vehicles experience, *ADFD*.

### B. First Come First Served (FCFS), First Depart First Served (FDFS)

Our baseline charging mechanism is FCFS. When there isn't sufficient power available to charge all of the vehicles that are currently plugged into chargers, we operate the switches to charge those vehicles that were plugged in first. We have considered two versions of FCFS, one uses information on the driving distance and the other doesn't.

The simple version charges vehicles until their battery is full. In the second version, once vehicles obtain the amount of power needed to reach their destination, they don't obtain additional power unless they can obtain that power without taking power from a vehicle that does not have sufficient power to reach its destination.

In the simulations we also consider FDFS, which is analogous to FCFS, but uses the expected departure time, rather than the arrival times to select the vehicles that receive charge. The vehicles that have the largest delay are charged first. If there aren't any vehicles that are delayed, or if all of the delayed vehicles are already selected in this round, the vehicles with smallest remaining time to receive charge are selected, to prevent them from becoming delayed.

### C. Round Robin (RR)

There are two versions of RR, one uses information on expected driving distances, and the other does not. The version that uses expected driving distances maintains two lists, $L_1$, with the vehicles that are plugged in and do not have sufficient power to reach their destinations, and a list, $L_2$, with the vehicles that are plugged in and have sufficient power to reach their destination, but do not have a full battery. The simple version maintains a single list, $L_1$, with all of the vehicles that are plugged in but do not have a full battery. List $L_2$ is empty. The number of vehicles in $L_1$ is $N_1$, and $L_2$ is $N_2$.

At the beginning of each 5 minute interval, we determine the number of vehicles $K$ that can be charged with the available power. If $K \leq N_1$, select the $K$ vehicles at the top of $L_1$, for charging and move them to the bottom of the list. If $N_1 < K \leq N_1 + N_2$, select the $N_1$ vehicles in $L_1$ and the $K$-$N_1$ vehicles at the top of $L_2$ for charging, move the selected vehicles from $L_2$ to the bottom of that list; If $N_1 + N_2 < K$, charge all of the vehicles.

RR gives the vehicles that are currently being charged, and are in the same list, the same amount of power on the average.

### D. Min-Max Based on the required Energy (MinmaxER)

The objective in MinmaxER is to minimize the maximum time until any charging vehicle can leave. This is equivalent to minimizing the maximum number of charging intervals for any of the vehicles to acquire sufficient power to leave. When we cannot charge all of the vehicles that are plugged in, MinmaxER selects those vehicles that require the largest number of charging intervals before they can leave.

A vehicle can reach its destination when the charge that it has acquired is:

$$E_R = (D/R) \times B \tag{2}$$

where $D$ is the required driving distance, $R$ is the range of a vehicle with a fully charged battery, and $B$ is the amount of energy in a fully charged battery. If the battery has charge $C$, and can acquire charge $E_5$ in a 5 minute interval, the number of intervals that the vehicle requires is:

$$N_I = (E_R - C)/E_5. \tag{3}$$

At the beginning of each 5 minute interval, we calculate $N_I$ for each vehicle that is plugged into a charger and select the vehicles with the largest values.

### E. Min-Max Delay Time (MinmaxDT)

The objective in MinmaxDT is to minimize the maximum time that any vehicle is delayed. If a vehicle does not have sufficient charge for its expected travel, it is delayed until it has acquired the charge. The delay is defined as the difference between the completion time of charging and the expected departure time.

We define a system as being min-max efficient when we distribute as much of the available power as possible, and when decreasing the delay of a vehicle, by selecting the vehicle for charging, does not increase the delay of a vehicle that will incur a larger delay. This definition is analogous to the definition of max-min fairness for the assignment of flows in a communications network [8,10].

In order to realize the objective, we construct an ordered list of the delay that vehicles will incur if they are charged during every available charging interval until they acquire sufficient power to reach their destinations. The delays can be positive or negative. We cannot completely schedule the vehicles' charging intervals because the number of vehicles that can be charged during any interval will change as the other loads change, and because other vehicles may arrive for charging.

The positive delays are the minimum delay that the vehicles will incur. The magnitude of a negative delay is the number of intervals in which a vehicle can be denied charge before incurring a positive delay. Charging the vehicles with the largest of the positive delays first minimizes the maximum of the positive delays. Selecting the vehicles with a smaller magnitude of negative delay before those with a larger magnitude will prevent a larger number of vehicles from incurring positive delays.

When there is sufficient power to charge all of the vehicles, they all receive charge. When there isn't sufficient power to charge all of the vehicles, we select the vehicles that will receive charge from the ordered list, until we have distributed all of the available power. Therefore, we distribute as much of the available power as possible.

In any interval in which a vehicle is not selected, its minimum delay increases by 1. The ordered list guarantees that vehicles whose delay is increased have a minimum delay that is less than or equal to the delay of the vehicles that are selected. None of the selected vehicles increases the delay of any vehicle that has a larger delay. Therefore, the system min-max efficient.

The number of charging intervals that a vehicle requires to reach the energy level at which it can depart, $N_I$, is calculated as in MinmaxER. Measuring time in 5 minute intervals, $T$ is the current time and $T_L$ is the expected time at which a vehicle would like to leave. If a vehicle is charged in every 5 minute interval, until it reaches its departure charge, its delay is:

$$D_T = N_I - (T_L - T) \tag{4}$$

The departure delay, $D_T$, is measured in 5 minute intervals.

Vehicles are ordered according to their values of $D_T$, and the vehicles with the largest values are selected first.

## IV. Evaluation

In this section, we describe the simulations and present the results. The simulations are based on the real world measurements presented in section II.

## A. Simulation

The simulations compare five charging mechanisms. FCFS uses the arrival time of the vehicles to charge those vehicles that are waiting the longest. RR uses the number of charging vehicles to share the available power equally. When a user wants to depart, if he has sufficient charge for his trip, he is not delayed. A delayed user departs as soon as he acquires sufficient charge. The FCFS and RR mechanisms can charge the vehicles until their batteries are full. However, the delays in the simulations are reduced by giving preference to those vehicles that do not have sufficient charge to depart, as explained in section III.B&C. Using this information provides a better comparison with the other rules, and provides a lower bound on delays incurred without using the information.

MinmaxER uses the expected driving distance, but not the expected departure time or the arrival time. It charges the vehicles with the greatest charge deficit first. FDFS uses the expected departure time to charge the vehicles that will be leaving the soonest or are delayed the most. FDFS also uses the expected driving distances, the same as FCFS or RR. And, MinMaxDT uses both the expected driving distances and the expected departure times, to minimize the maximum delay.

The simulations are conducted for around 1500 vehicles, as explained in section II. The results are reported as a function of the supply-to-demand ratio, *SDR*, the total power available to charge vehicles over the total power required by the vehicles. The value of *SDR* ranges from 1 to 3. For each curve *SDR* increases in steps of 0.05 from 1 to 1.2, 0.2 from 1.2 to 2, and 1 from 2 to 3. For each point on each on each curve, the simulations are run for 15 days. The measurements are only taken for vehicles that arrive after the 4$^{th}$ day, to allow the system to become stable, and until the 13$^{th}$ day, so that all vehicles included in the statistics have completed charging.

The time required to charge an electric vehicle is significant. A vehicle that arrives with an empty battery and requires a full battery takes more than 16 hours to obtain the charge from a 110 volt, 15 amp circuit. About 5% of the vehicles in our simulation cannot leave by their expected departure time, even if they are charged continuously. We are interested in the delays caused by power short falls, rather than the unrealistic expectations of the users. When the expected departure time, selected from the distributions, is less than the minimum departure time required when the vehicle is charged continuously, the departure time is increased to the larger value.

## B. Results

Fig.1, Fig.2, and Fig.3 present the results for the five charging schemes using 110 volt, 15 amp circuits. Fig.1 is the fraction of the vehicles that are delayed as a function of the *SDR*, and Fig.2 is the average delay incurred by the vehicles that are delayed. Fig.3 is the distribution of delays when *SDR*=1.2.

From Fig.1, we see that MinmaxDT outperforms the other mechanisms by such a large amount that it the clear choice, if it is at all possible to obtain accurate information about expected driving distances and departure times. With MinmaxDT, almost none of the vehicles are delayed when *SDR* is as small as 1.05, the available power over a 24 hour period exceeds the demand by as little as 5%. With FDFS, which also requires expected departure times, *SDR* must be greater than 1.5 to guarantee that less than 5% of the vehicles are delayed. Since MinmaxDT out performs FDFS, we will not consider using FDFS.

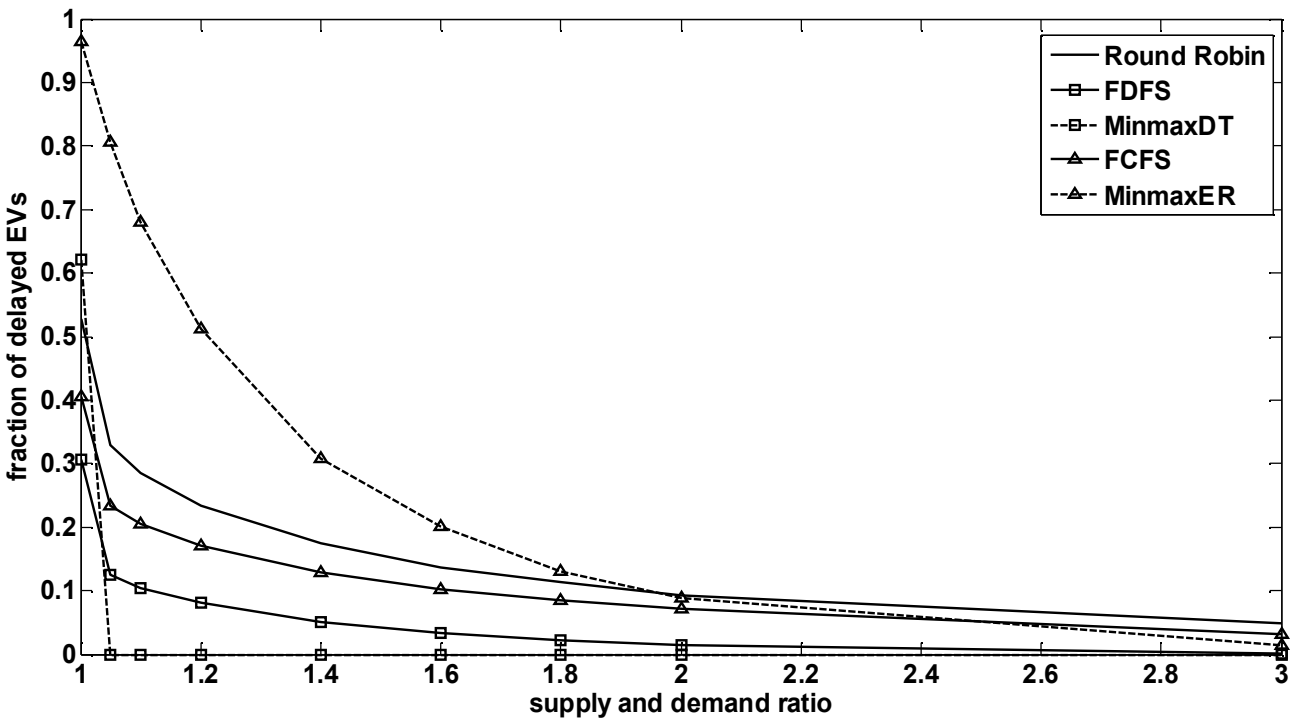

Fig. 1. Fraction of delayed vehicles

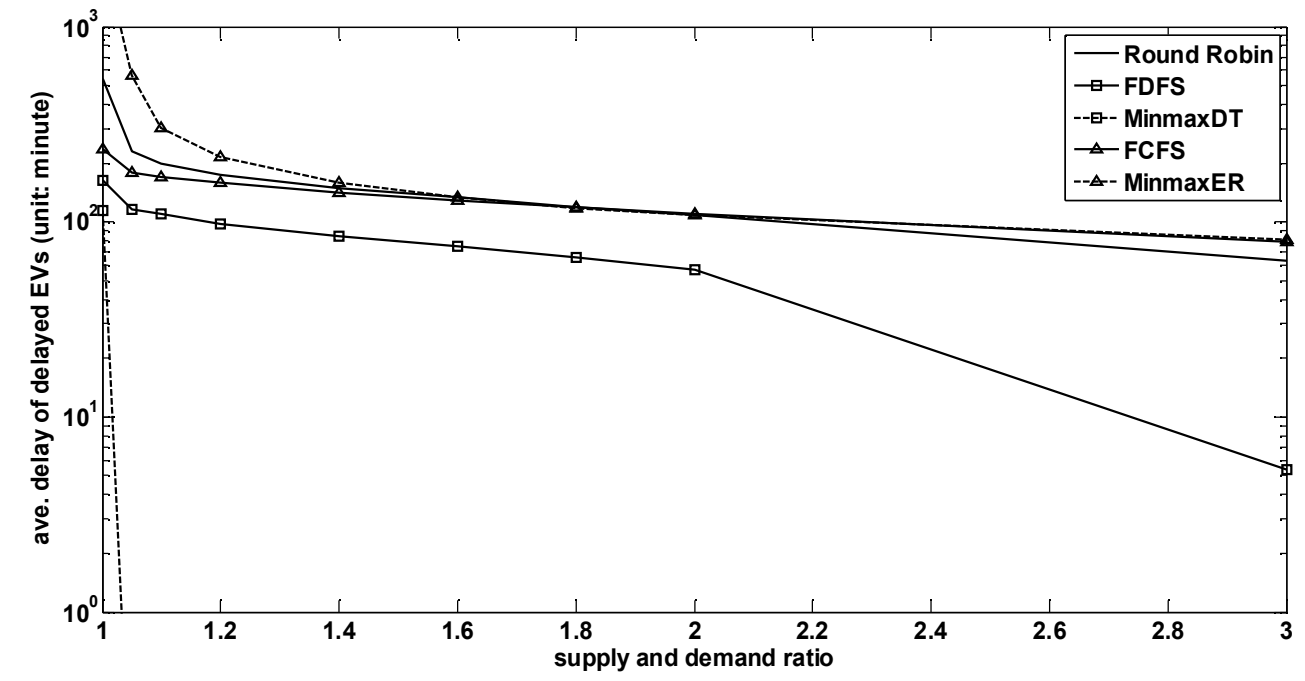

Fig. 2. Average delay of delayed vehicles

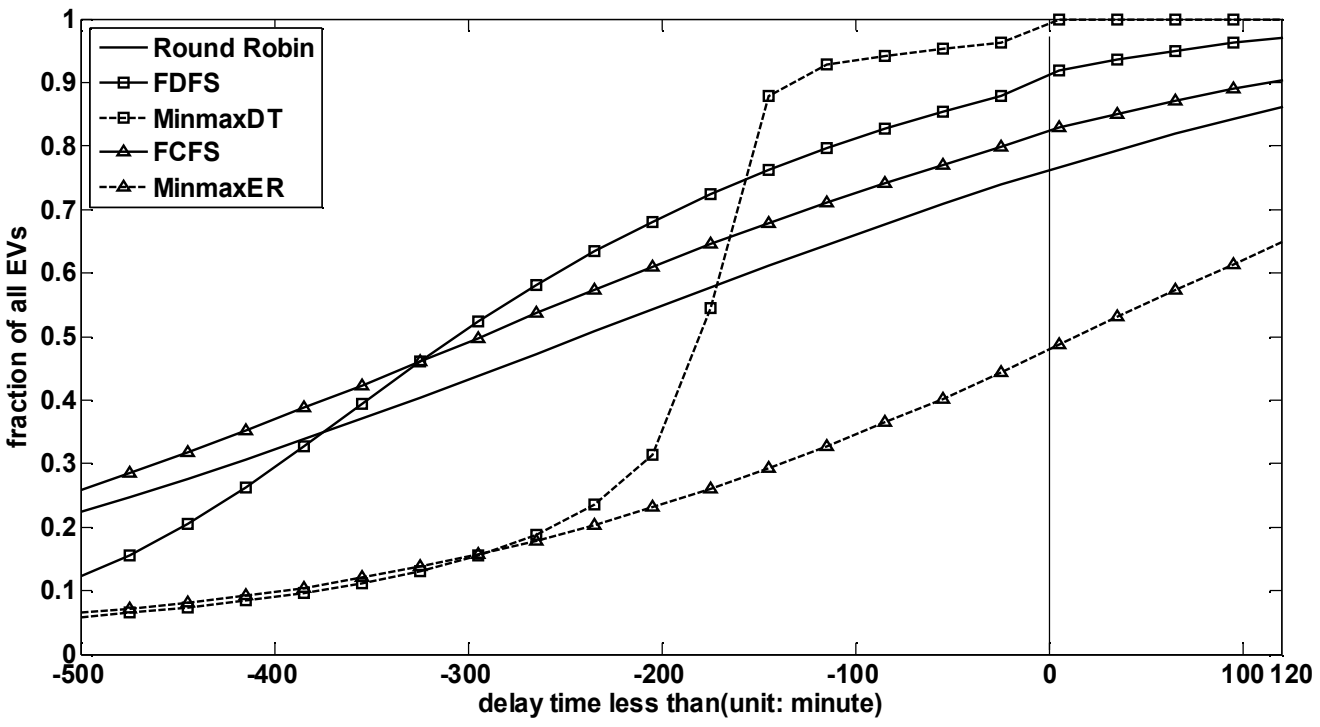

Fig. 3. The delay distribution at SDR=1.2

The other three schemes, which do not require expected departure times, require *SDR*>2 for less than 10% of the vehicles to be delayed. More importantly, from Fig.2, the average delay of those vehicles that are delayed can exceed two hours. From Fig.3, at *SDR* =1.2, more than 10% of the delayed vehicles using FCFS or RR, and 35% of the vehicles using MinmaxER, have average delays exceeding 2 hours. When *SDR* is increased to 3, 3% or 4% of the delayed vehicles using FCFS or RR still have average delays of about 1½ hours. Most of the delayed vehicles depart in the morning, when the power available for charging is decreasing. As a result the vehicles that are delayed are delayed for a longer time. These delays are not tolerable, and will restrict the use of electric vehicles to those with shorter commutes, who are less likely to suffer these delays.

We note that MinmaxER is outperformed by all of the other schemes. A larger fraction of the vehicles are delayed and their delays are larger. The strategy for MinmaxER is flawed, and the scheme should never be used. As a vehicle in MinmaxER approaches the charge that it requires to reach its destination, all newly arrived vehicles will requires more charge and take precedence. As a result the closer a vehicle gets to achieving its required charge, the longer it waits to acquire that charge.

As a final observation, we note that while FCFS and RR are close in performance, However, FCFS outperforms RR. We conclude that it is more important to complete the charging of those vehicles that have been waiting the longest, than it is to distribute the charge fairly among all of the vehicles that are currently charging. If we do not use information provided by a user, we should use FCFS.

## V. Conclusion

A smart grid can control which electric vehicles are charged by a remotely controlled switch at the charging station. In this paper, several fairness algorithms that are based on information that can be measured, estimated, or requested from a user, are presented to control the average rate that each vehicle receives charge. They are more precise mechanisms to control power distribution than pricing mechanisms that control demand by encouraging customers to shift their use to less expensive, low use periods of the day. In addition, the mechanism does not favor the wealthy, who are less likely to allow a change in price to delay the use of their vehicle.

We have compared five techniques that collect and use different types of information to select the vehicles that receive charge during successive 5-minute intervals. The techniques are applied to vehicles that are charged at home for a period of 6 to 22 hours. We vary *SDR*, the ratio of the total power available to charge the vehicles to the total power that they require, and measure the fraction of the vehicles that cannot receive sufficient power before they would like to leave, and the delays that they incur.

Two of the mechanisms, FCFS and RR, can be implemented without any estimated information. FCFS outperforms RR. It delays fewer vehicles. With FCFS, if we increase the charging stations from 110 volt, 15 amp circuits, to 220 volt, 30 amp stations, at *SDR*=2 the fraction of the vehicles that are delayed is reduced from 10% to about 2%. With higher power charging stations, an FCFS selection strategy provides acceptable delays.

The other three mechanisms use the expected departure time and expected driving distance to control the selection process. One of the techniques, MinmaxDT, based upon minimizing the maximum delay, is vastly superior to any of the four other techniques considered. When the power available during the course of the day exceeds the power that is required by vehicles by as little as 5%, almost none of the vehicles are delayed. These results are obtained using 110 volt, 15 amp circuits.

The advantages of using MinmaxDT make it worthwhile to consider how we might obtain accurate information about the departure and driving distance. One possibility is to estimate a user's driving habits [11], allow the user to enter different values for special circumstances, and provide penalties for entering misleading information. Most drivers leave home at about the same time each day and drive nearly similar distances. We can estimate the next departure time and driving distance as a running average of the past days, probably scheduling the departure ½ hour earlier and adding 10 mile to the distance, for emergencies. The estimates can be displayed for the user, and he can be allowed to change the estimates when he has different needs. If his entered value of the departure time is much earlier than the actual departure time, or if the entered value of the driving distance is much greater than the distance that he drives, he can be penalized the next time he makes an entry by subtracting mileage or adding time to his entry.

## Acknowledgment

This work was supported by the National Natural Science Foundation of China (No. 91338107), the Fundamental Research Funds for the Central Universities (No. 2014SCU11013). We would like to thank Yasser Mohammed to proofread this paper.

## References

[1] P. Samadi, A. Mohsenian-Rad, R. Schober, and etc,"Optimal Real-Time Pricing Algorithm Based on Utility Maximization for Smart Grid," IEEE SmartGridComm Conf., pp.415-420, Oct. 2010.

[2] N. Rotering and M. Ilic "Optimal charge control of plug-in hybrid electric vehicles in deregulated electricity markets", IEEE Trans. Power Syst., Vol. 26, No. 3, pp.1021-1029, 2011

[3] R. A. Verzijlbergh, M. O. W. Grond, Z. Lukszo, J. G. Slootweg, and M. D. Ilić, "Network Impacts and Cost Savings of Controlled EV Charging",IEEE Trans. on Smart Grid,Vol.3, No.3, pp.1203-1212, 2012

[4] S. Tang, C. Li, P. Zhang, Y. Tan, Z. Zhang and J.Li, "An Optimized EV Charging Model Considering TOU Price and SOC Curve", IEEE Transactions on Smart Grid, vol.3, no.1, pp. 388-393, 2012

[5] Brian McKenzie, Melanie Rapino, "Commuting in the United States: 2009", American Community Survey Reports, Sep.2011.

[6] www.bts.gov/publications/omnistats/volume_03_issue_04/html

[7] www1.eere.energy.gov/cleancities/toolbox/pdfs/driving_san_diego.pdf

[8] D. Bertsekas, R. G. Gallager, Data Networks, Prentice-Hall Inc, 1992.

[9] James R. Slagle and Henry J. Hamburger, "An expert system for a resource allocation problem", Communications of the ACM, Vol.28, No. 9, pp. 994-1004, 1985

[10] Tzeng Hong-Yi, Sin Kai-Yeung, "On max-min fair congestion control for multicast ABR service in ATM", IEEE Jour. Selected Areas in Commun., vol.15, no.3, pp.545-556, Apr. 1997.

[11] Y. Zhou, N. Maxemchuk, X. Qian and Y. Mohammed, "A Weighted Fair Queuing Algorithm for Electric Vehicle Charging Management on a Smart Grid", in Proc. of IEEE Online Conference on Green Communications (GreenCom), pp.132-136, Oct. 2013.